\documentclass[11pt]{article}
\usepackage{moriond,epsfig}
\def\jpsi{J/$\psi$}
\def\psip{$\psi^\prime$}
\begin{document}

\vspace*{4cm}
\title{FIRST RESULTS from the NA60 EXPERIMENT at CERN}
\author{G. Usai$^3$ for the NA60 Collaboration: \\  
R.~Arnaldi$^{11}$, 
K.~Banicz$^4$,
K.~Borer$^1$,
J.~Buytaert$^4$,
J.~Castor$^5$,
B.~Chaurand$^8$,
W.~Chen$^2$,  
B.~Cheynis$^7$,  
C.~Cicalo$^3$,  
A.~Colla$^{11}$, 
P.~Cortese$^{11}$,  
A.~David$^6$, 
A.~Devaux$^5$,  
A.~Drees$^{10}$, 
L.~Ducroux$^7$, 
H.~En'yo$^9$,  
A.~de~Falco$^3$,  
A.~Ferretti$^{11}$,  
M.~Floris$^3$, 
P.~Force$^5$, 
A.~Grigorian$^{12}$,  
J.-Y.~Grossiord$^7$,  
N.~Guettet$^5$,  
A.~Guichard$^7$,
H.~Gulkanian$^{12}$,  
J.~Heuser$^9$,  
M.~Keil$^4$,  
L.~Kluberg$^8$, 
Z.~Li$^2$,  
C.~Louren\c{c}o$^4$, 
J.~Lozano$^6$,  
F.~Manso$^5$, 
N.~de~Marco$^{11}$,
A.~Masoni$^3$, 
A.~Neves$^6$,  
H.~Ohnishi$^9$, 
C.~Oppedisano$^{11}$,
G.~Puddu$^3$, 
E.~Radermacher$^4$, 
P.~Rosinsky$^4$,  
E.~Scomparin$^{11}$,  
J.~Seixas$^6$,  
S.~Serci$^3$, 
R.~Shahoyan$^6$, 
P.~Sonderegger$^6$,  
G.~Usai$^3$,  
H.~Vardanyan$^{12}$, 
and H.~W\"ohri$^4$}

\address{$^1$ 
Laboratory of High Energy Physics, University of Bern, Bern, Switzerland\\
$^2$ Brookhaven National Laboratory, New York, USA \\
$^3$ Universit\`a di Cagliari and INFN, Cagliari, Italy \\
$^4$ European Laboratory for Particle Physics, Geneva, Switzerland \\
$^5$ LPC, Universit\'e Blaise Pascal and CNRS-IN2P3, Clermont-Ferrand, France \\
$^6$ Instituto Superior T\'ecnico, Lisbon, Portugal \\
$^7$ IPN, Universit\'e Claude Bernard Lyon-I and CNRS-IN2P3, Lyon, France \\
$^8$ LLR, Ecole Polytechnique, CNRS-IN2P3, Palaiseau, France \\
$^9$ RIKEN, Radiation Laboratory, Japan \\
$^{10}$ SUNY Stony Brook, New York, USA \\
$^{11}$ Universit\`{a} di Torino and INFN, Turin, Italy \\
$^{12}$ YerPhI, Yerevan Physics Institute, Yerevan, Armenia}

\maketitle

\abstract{Since 1986, several heavy ion experiments have studied some
signatures of the formation of the quark-gluon plasma and a few
exciting results have been found.  However, some important questions
are still unanswered and require new measurements.  The NA60
experiment, with a new detector concept that vastly improves dimuon
detection in proton-nucleus and heavy-ion collisions, studies several
of those open questions, including the production of open charm.  This
paper presents the experiment and some first results from data
collected in 2002.}

\bigskip

In the study of the possible formation of a deconfined state of
strongly interacting matter, where chiral symmetry is restored, by
colliding heavy nuclei at high energies, a few very interesting
observations have been made.  However, certain aspects in the
interpretation of these measurements remain unclear and require a
better look.  The NA60 experiment addresses some of these issues,
namely those that can be studied through dilepton production.

The dimuon continuum in the mass window between the $\phi$ and the
\jpsi\ resonances is very well described in proton-nucleus collisions
by simply adding the Drell-Yan dimuons to the expected yield of muon
pairs from simultaneous semi-muonic decays of D meson pairs.  In S-U
and Pb-Pb collisions, however, the data exhibits an important excess
with respect to these expected sources~\cite{NA50imr}.  This excess
may be due to a rather strong enhancement of charm production (up to a
factor 3 in central Pb-Pb collisions) or to thermal dimuons produced
in the quark gluon plasma phase.  NA60 will distinguish these two
possibilities and directly measure the open charm yield, an important
measurement in itself, being the heaviest flavour that can be produced
today in heavy-ion collisions and given the significant enhancements
observed in the (multi-)strangeness sector.

So far, charm has only been studied through \jpsi\ and \psip\
production, $c\bar{c}$ bound states, namely in the NA38 and NA50 SPS
experiments.  The well-known \jpsi\ ``anomalous
suppression''~\cite{NA50psi} is generally considered to be one of the
most interesting observations done so far in the context of the search
for the formation of the quark-gluon plasma phase.  However, its
detailed interpretation poses some difficulties.  For instance, it is
likely that what is being suppressed is the $\chi_c$ meson,
responsible for around 30--40\,\% of the \jpsi\ mesons measured in
elementary pp collisions.  It is also very strange that this
($\chi_c$) suppression happens at rather high energy densities, higher
than 2~GeV/fm$^3$, while the value expected from lattice QCD
calculations is $\epsilon_c\sim 0.7$~GeV/fm$^3$.  Maybe the energy
density is not the variable driving charmonium suppression.  To
clarify these questions, NA60 will complement the S-U (NA38) and the
Pb-Pb (NA50) data with a detailed study of In-In collisions.

Also the CERES experiment observed a very interesting excess of
dileptons in the mass window around 500~MeV, equally seen in S-Au and
in Pb-Au collisions~\cite{CERES}.  It is tempting to associate these
observations with the predicted restoration of chiral symmetry in the
hot and dense medium created in these collisions.  NA60 will provide a
new measurement, using dimuons instead of dielectrons, with completely
independent systematic uncertainties and good statistics, mass
resolution and signal to background ratio.  Surely, the new
measurements will be very helpful in clarifying the interpretation of
these results.


The NA60 experiment complements the muon spectrometer and zero-degree
calorimeter (ZDC) previously used in NA50 with a completely new
detector concept in what concerns the target region.  The basic idea
is to place an ``eye'' in the vertex region, composed of
state-of-the-art silicon detectors that track the beam and the
particles produced in the target, within the angular acceptance of the
muon spectrometer~\cite{NA60proposal}.

The beam tracker is composed of two silicon microstrip stations,
placed 20~cm apart, just upstream of the target.  The sensors have 24
strips of 50~$\mu$m pitch, allowing to infere the \emph{transverse}
coordinates of the interaction point, at the target, with a resolution
of 20~$\mu$m.  Being operated at 130~K, in vacuum, we have seen (in
year 2000) that these sensors continue to be sensitive to the beam
ions even after more than 30~days of continuous running, with beam
intensities of $7\times 10^7$ Pb ions per burst.  At room temperature,
we would not expect the sensors to survive the first day.  The very
fast signals (rise time less than 500~ps) together with a devoted (and
fast) multi-hit time recorder system, ensure a time accuracy of 1.7~ns
and a double-peak resolution better than 10~ns, useful to reject beam
pile-up.

After the target, and inside a 2.5~T dipole field, sits the silicon
tracking telescope, that tracks the charged particles produced within
the angular acceptance of the muon spectro\-me\-ter.  For the ion
runs, the telescope is made of silicon pixel detectors, to cope with
the very high particle multiplicities produced.  In the proton runs we
can also use microstrip planes, to extend the tracking distance up to
40~cm from the target, with much less material budget than with pixel
assemblies, and using a faster clock (40~MHz), relevant to fight
interaction pile-up at the very high proton beam intensities (more
than $10^9$~protons per 5~s burst).

\begin{figure}[t]
\centering
\begin{minipage}[t]{0.51\textwidth}
\resizebox{\textwidth}{!}{\includegraphics*{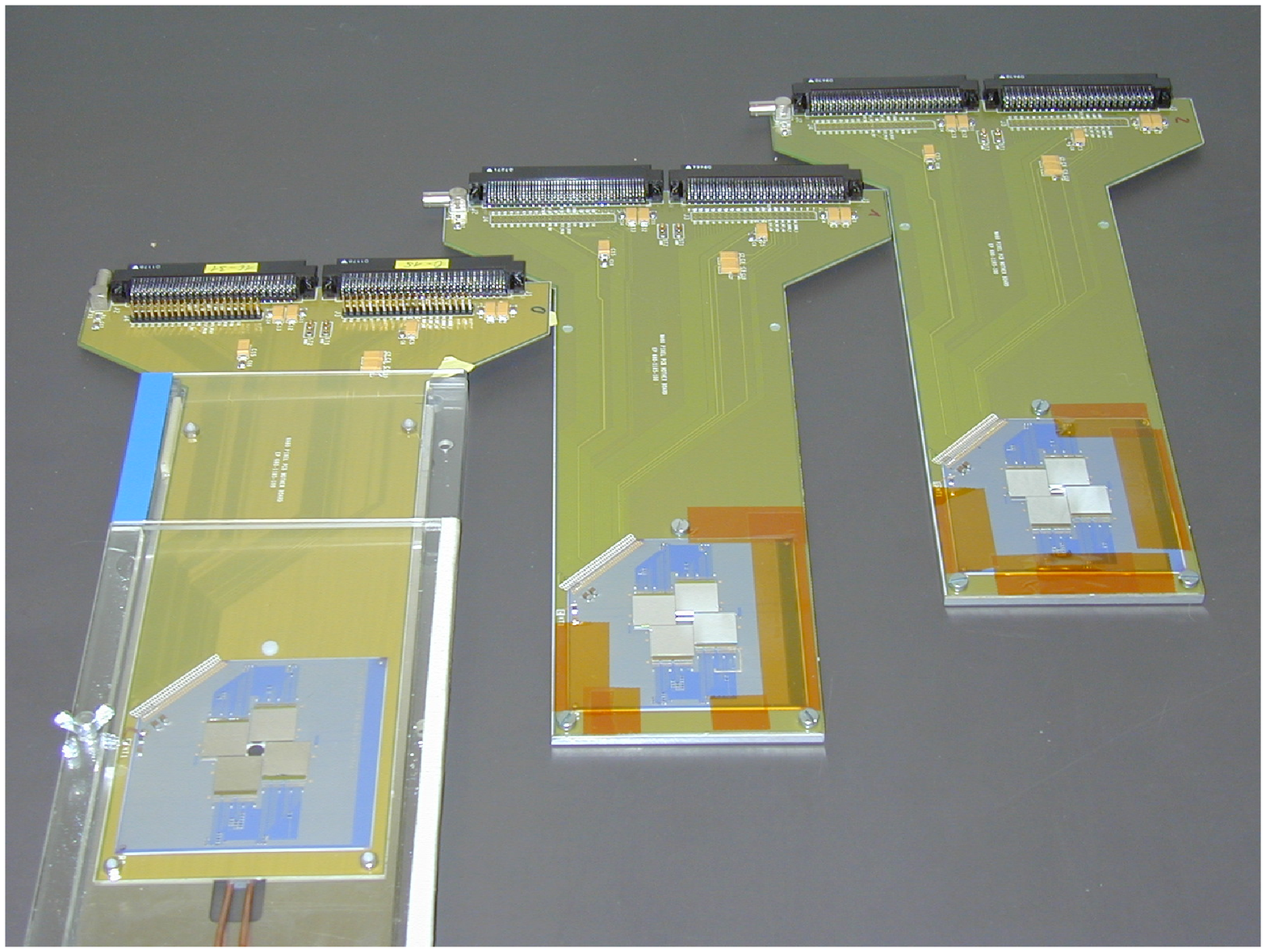}}
\caption{4-chip planes used in October 2002.}
\label{vertel}
\end{minipage}
\hfill
\begin{minipage}[t]{0.45\textwidth}
\resizebox{\textwidth}{!}{\includegraphics*{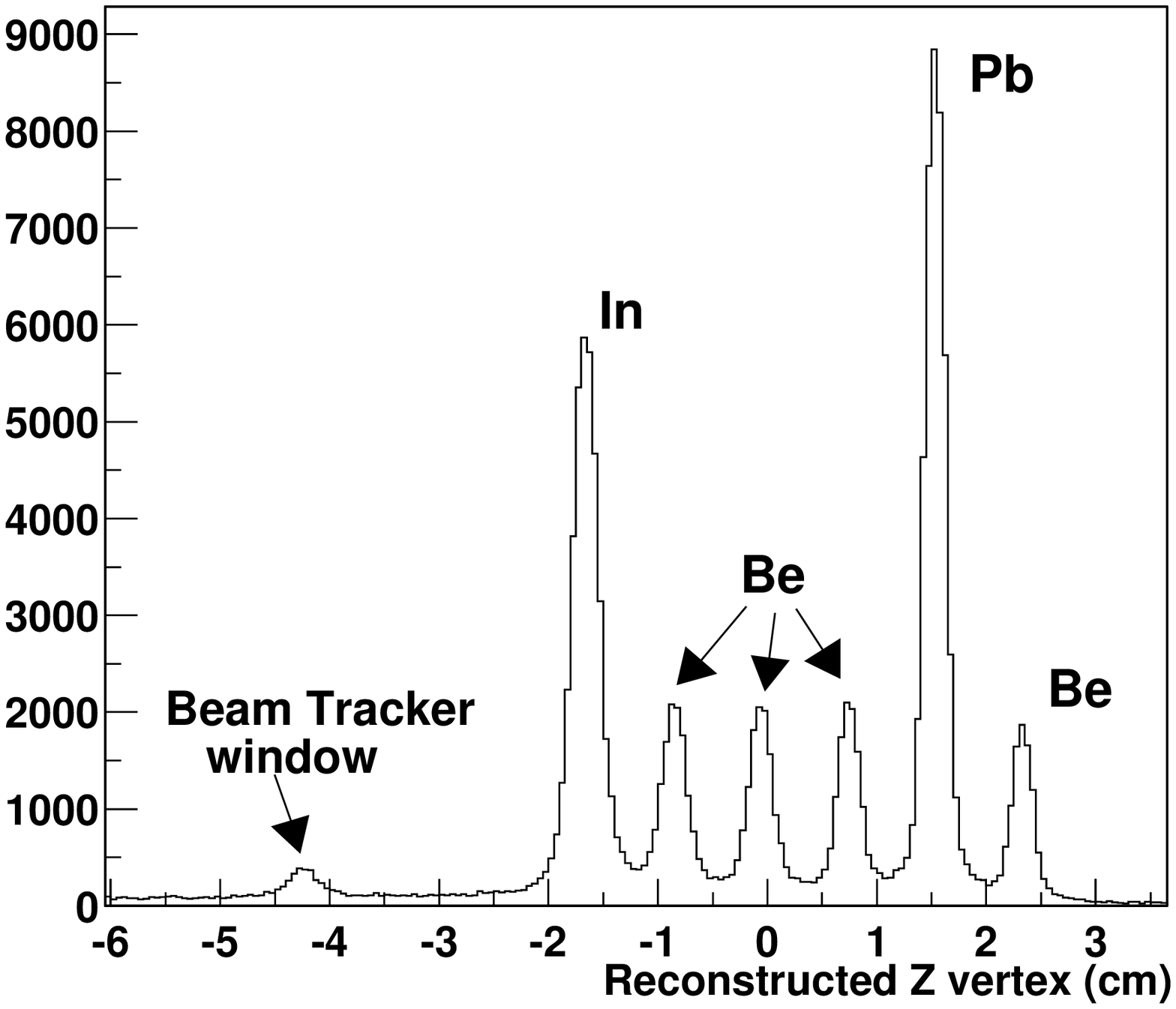}}
\caption{$Z$-vertex distribution for the p-A run.}
\label{pazvertex}
\end{minipage}
\end{figure}

The NA60 silicon pixel telescope will consist of eight 4-chip pixel
planes (illustrated in Fig.~\ref{vertel}) followed by 4 large
stations, each made of two 8-chip planes, to cover the solid angle
defined by the muon spectrometer.  The read-out pixel chip is a matrix
of $256 \times 32$ pixels, each cell being $50\times425~\mu$m$^2$.
These chips were designed for the ALICE and LHC-B
experiments~\cite{pixelchip}, work at 10~MHz and are radiation hard up
to around 30~Mrad.  They are bump-bonded to 300~$\mu$m thick silicon
sensors and assembled in ceramic hybrids.  The DAQ system reads the
almost 1 million pixel channels through a PCI-based read-out
electronics system.

The silicon vertex telescope tracks the charged particles and selects
those that provide the best match to the muons identified and measured
in the muon spectrometer.  The matching allows us to access the angles
and momenta of the muons before they suffer the multiple scattering
and energy loss induced by the absorbers, thereby significantly
improving the dimuon mass resolution.  Since the matching probability
is lower for muons from K and $\pi$ decays, the signal to background
ratio is also improved.  Finally, measuring the offset in the
transverse plane between the muon's trajectory and the interaction
point allows us to distinguish an event sample dominated by muons
resulting from decays of charmed mesons from an event sample composed
of prompt dimuons.

\begin{figure}[ht]
\centering
\begin{minipage}[t]{0.485\textwidth}
\resizebox{\textwidth}{!}{\includegraphics*{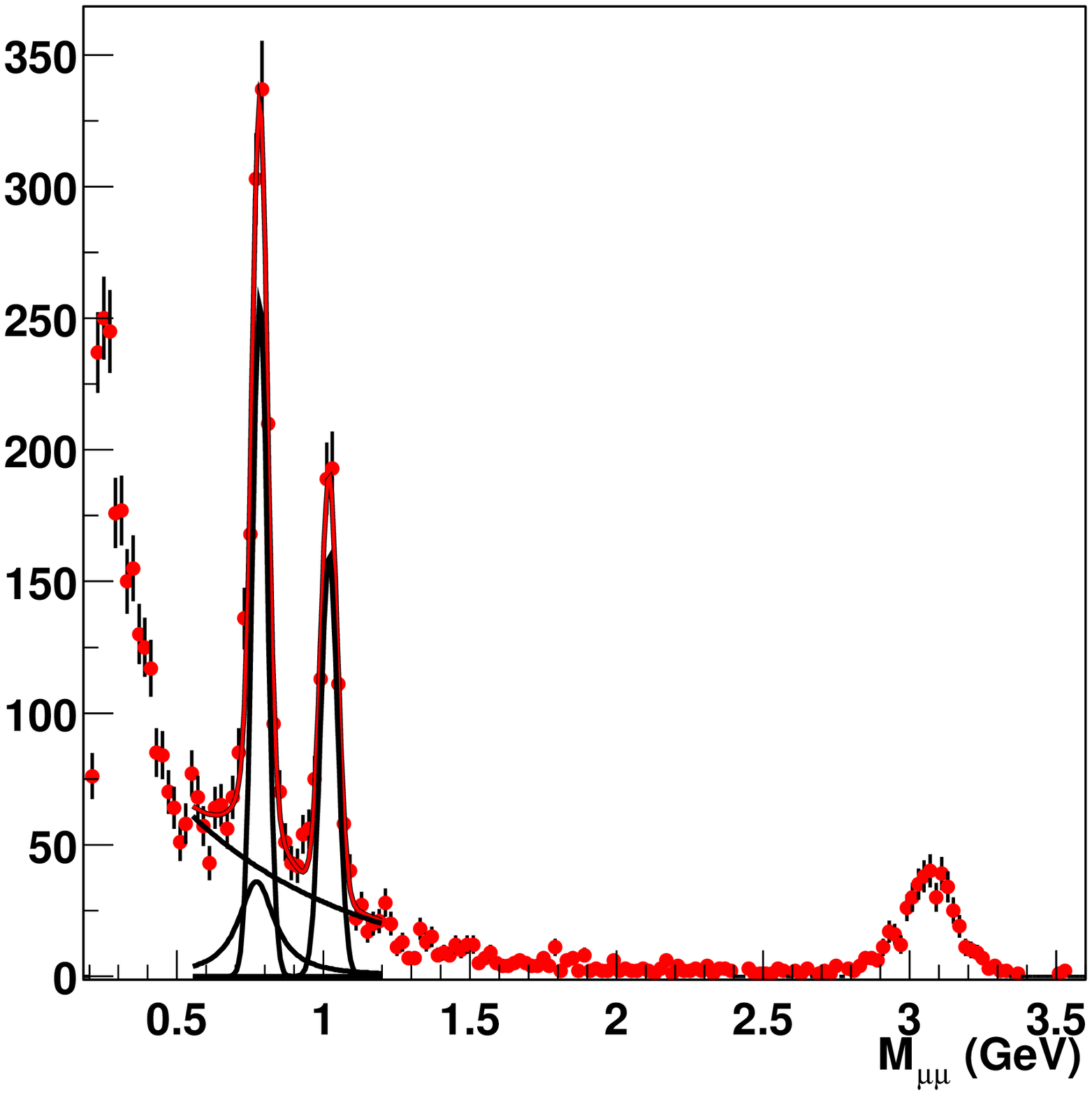}}
\caption{Dimuon mass distribution for p-Pb collisions}
\label{ppbmass}
\end{minipage}
\hfill
\begin{minipage}[t]{0.47\textwidth}
\resizebox{\textwidth}{!}{\includegraphics*{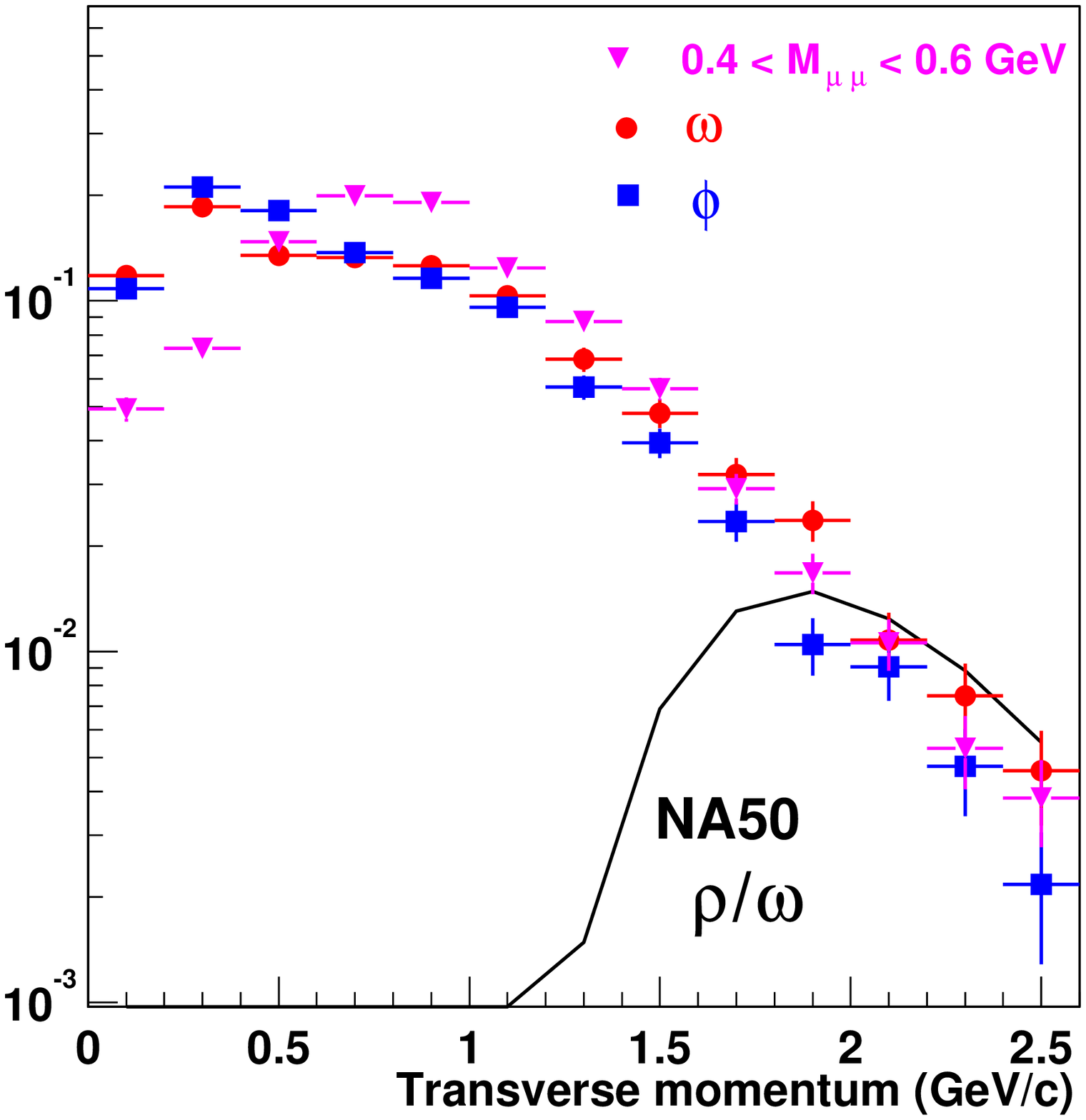}}
\caption{$p_{\rm T}$ distributions measured for different mass regions.}
\label{pt}
\end{minipage}
\end{figure}

During around two weeks in June 2002, NA60 collected proton-nucleus
data, using a dimuon trigger and a 400~GeV proton beam incident on Be,
In and Pb targets.  The targets (6 in total) were placed every 8~mm
and were 2~mm thick, while the interaction point was reconstructed
from the tracking of the charged particles in the silicon telescope
with a resolution of $\sim$\,900~$\mu$m, in the beam direction (see
Fig.~\ref{pazvertex}).  The dimuon mass, after matching, is measured
with a resolution around 25~MeV at the $\omega$ mass and around 30~MeV
at the $\phi$ mass (see Fig.~\ref{ppbmass}), a major improvement with
respect to NA50.  Furthermore, the acceptance for low mass dimuons
extends now down to very low $p_{\rm T}$ (see Fig.~\ref{pt}), mostly
thanks to the presence of the dipole magnet in the target region,
which bends into the acceptance of the muon spectrometer tracks that
would otherwise be lost in the beam dump.  This should allow NA60 to
perform measurements complementary to CERES and, also, should provide
a $p_{\rm T}$ distribution of the $\phi$ that overlaps with the
measurements of NA49, contrary to NA50 data.

\begin{figure}[ht]
\begin{minipage}[t]{0.48\textwidth}
\resizebox{\textwidth}{!}{\includegraphics*{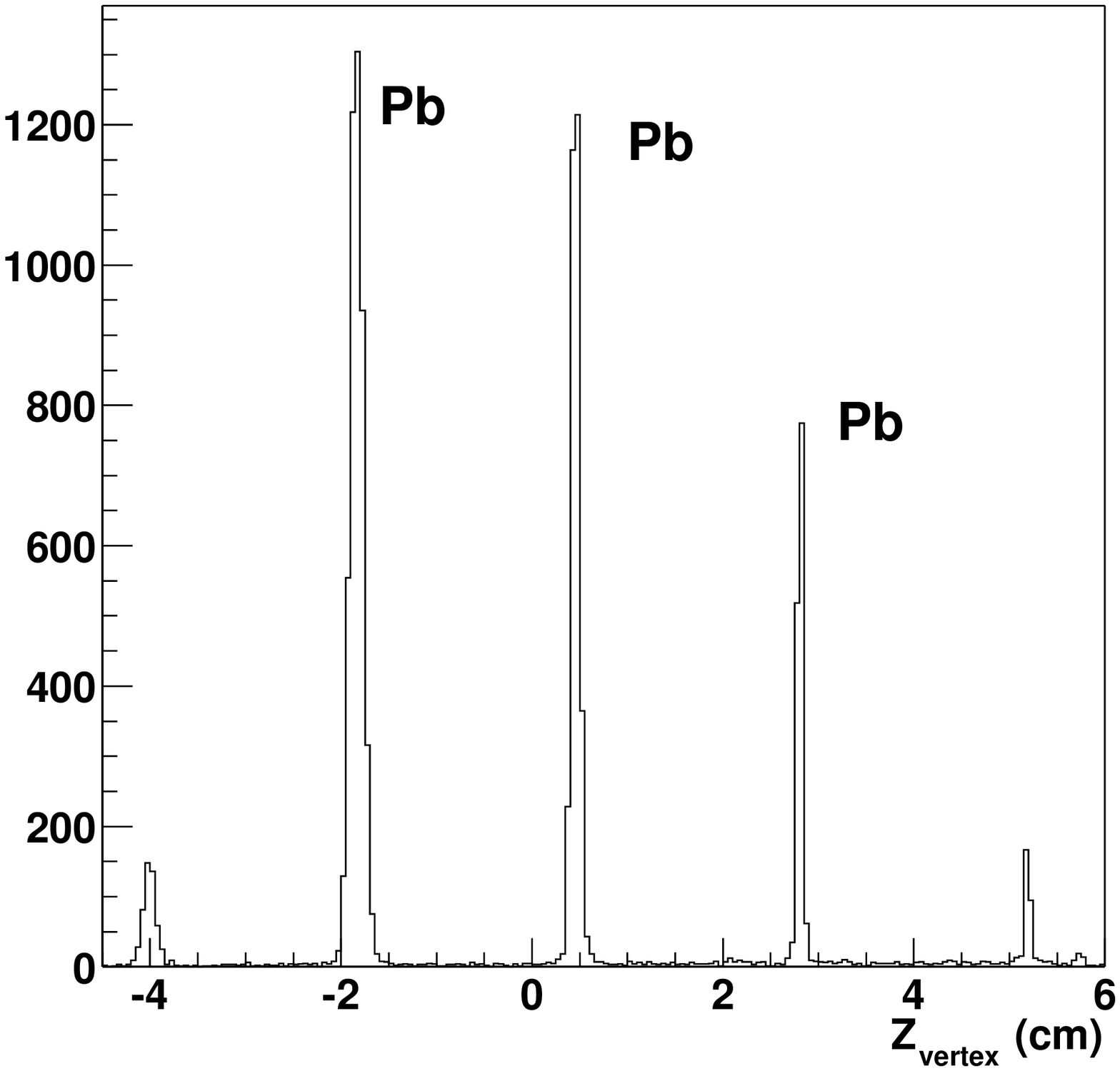}}
\caption{$Z$-vertex distribution measured in Pb-Pb collisions, during
  the 30~GeV per nucleon run.}
\label{pbzvertex}
\end{minipage}
\hfill
\begin{minipage}[t]{0.485\textwidth}
\resizebox{\textwidth}{!}{\includegraphics*{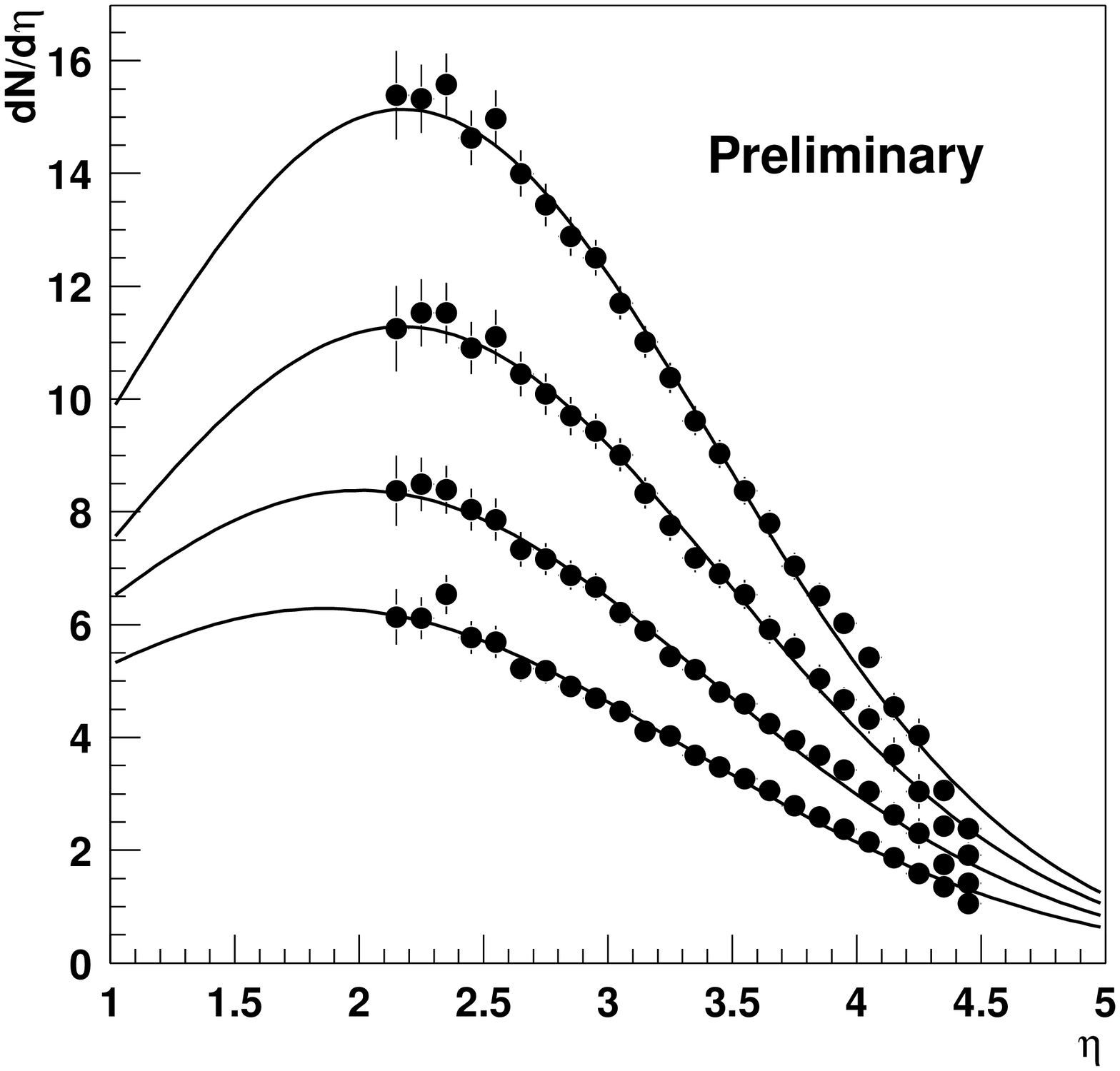}}
\caption{Rapidity densities of charged particles measured for
  different centrality ranges.}
\label{ydensities}
\end{minipage}
\end{figure}

In October 2002, NA60 had a test run with a Lead beam of 30 and 20~GeV
per nucleon (5 days each), incident on three Lead targets of different
thicknesses: 1.5, 1.0 and 0.5~mm, along the beam.  Three pixel planes
were successfully operated and were able to track the several tens of
charged particles produced in their angular acceptance.  In spite of
the small number of tracking planes, the vertex of the Pb-Pb collision
could be reconstructed with a resolution better than 200~$\mu$m along
the beam direction (see Fig.~\ref{pbzvertex}), and of around 20~$\mu$m
for the $X$ transverse coordinate (all the three planes had the
50~$\mu$m side of the cells along the $X$ coordinate), as expected
from the physics performance simulations.

As a by product of the Pb test run, we measured the rapidity densities
of the charged particles produced in the 30~GeV per nucleon Pb-Pb
collisions, for several centrality classes estimated through the ZDC,
as can be seen in Fig.~\ref{ydensities}.  For this measurement, the
three pixel planes were placed as close as possible to the targets, so
that the angular coverage would include mid-rapidity (2.08 at an
incident energy of only 30~GeV/nucleon), and the multiplicity was
estimated through a cluster counting procedure~\cite{SQM03}.
Acceptances and efficiencies were considered for each target and each
pixel plane.  The spurious contributions from secondary interactions
and from $\delta$-rays produced by the incident ions were estimated
through Monte Carlo simulation, using UrQMD and GEANT.

\bigskip

In conclusion, NA60 has taken first data in year 2002, with proton and
Pb ion beams, with a new silicon vertex telescope in the target
region.  With the proton runs we have collected dimuon data showing
good mass resolution and improved acceptance for low mass and $p_{\rm
T}$ dimuons.  In the ion run we measured the interaction point with
excelent accuracy.  These results confirm the feasibility of the
experiment and give good perspectives for the next runs with proton
and Indium beams.

\end{document}